\documentclass[preprint,aps,eqsecnum]{revtex4}
\usepackage{graphicx}
\usepackage{latexsym}
\begin{document}
%\draft
%\preprint{\vbox{\baselineskip=12pt}
%\rightline{}
%\rightline{hep-th/yymmnnn}
%\title{de Sitter branes with bulk tachyon matter}
\title{de Sitter branes with a bulk scalar}
\author{Supratik Pal \footnote{Electronic address: {\em
      supipal@rediffmail.com ; supratik\_v@isical.ac.in}}}
${}^{}$
\affiliation{Inter--University Centre for Astronomy and Astrophysics, 
Post Bag 4, Ganeshkhind, Pune 411 007, India, and \\
Physics and Applied Mathematics Unit, Indian Statistical Institute,
203 B.T. Road, Kolkata 700 108, India}
\author{Sayan Kar \footnote{Electronic address: {\em sayan@cts.iitkgp.ernet.in}}}
${}^{}$
\affiliation{Department of Physics and Centre for
Theoretical Studies, Indian Institute of Technology Kharagpur 721 302, India}
\vspace{1in}

\begin{abstract}

We propose new braneworld models arising from a scalar field in the bulk.
In these examples, the induced on--brane line element is de Sitter (or anti de 
Sitter) and the bulk (five dimensional) Einstein equations can be exactly 
solved to obtain warped spacetimes. The solutions thus derived are single 
and two-brane
models -- one with {\em thin} branes while the other one of the
{\em thick} variety. The  field profiles and the potentials are obtained
and analysed for each case. We note that
for the {\em thick} brane scenario the field profile resembles a kink, whereas
for one or more {\em thin} branes, it is finite and bounded in the domain
of the extra dimension. 
We have also addressed the localisation  of gravity and other matter fields on 
the brane for these braneworld models.

\end{abstract}

%\pacs{}

\maketitle

\section{Introduction}

Theories with extra dimensions have been around for quite some time \cite{gsw}.
A comparatively new and qualitatively different idea is the warped braneworld 
scenario, which has, of late, shown much promise in solving some fundamental
problems (eg. the hierarchy problem) in high energy physics.
In this  scenario, the observed four dimensional
universe is assumed to be embedded in a higher dimensional background
and is warped by a function of the extra dimension -- the so-called `warp factor'.
The simplest realisation of such warped braneworld models is the Randall-Sundrum (RS)
thin brane model \cite{rs}, where the brane is mathematically represented
by a delta function singularity in the higher dimensional spacetime. 
Subsequently, different thin and thick brane models \cite{local, keha, eva, sgnum, sg, giov1, giov2,
phan, increasing, tach, giov3, asym, 6dloc, 6dten, 6d, rksk6d}
have been constructed using the five (or higher) dimensional Einstein-scalar 
set of equations, with the
bulk being constituted of either a cosmological constant or a scalar field, or both.
This gives rise to a spectrum of different possible warped higher dimensional 
spacetimes.  In most of the models, the brane is flat, \textit{i.e.},
the induced metric on the brane is scaled Minkowski. Later, 
a few braneworld models considered the curvature of the embedded brane.
Some of them investigated the scenario with an induced cosmological metric on 
the brane embedded in a higher dimensional spacetime
using transformed coordinates  \cite{ds, ds2, cosmo, bowcock, kanticos, mukoh}. 
However, in most of these braneworld models for which a
FRW geometry on the brane is recovered,  the bulk metric is given by either a 
Schwarzschild-anti-de-Sitter
(SchAdS) or a Vaidya-anti-de-Sitter (VAdS) black hole \cite{maartrev, langrev}, depending upon whether 
the bulk consists of only a cosmological constant or a radiative field.
It is only when the  mass of the bulk black hole vanishes, it leads to a warped
geometry after coordinate transformation. What turns out is that
none of the braneworld models compatible to a FRW metric on the brane, could, 
actually, provide an exact warped geometry.  
In this article, our basic goal is to propose exact, warped braneworld models 
%arising from a {\em tachyon matter} field in the bulk, 
giving rise to a cosmological metric on the brane.
Our model, in a sense, falls in the class of exact higher
dimensional warped spacetimes where the on-brane metric is FRW. 
We shall show, in due course, that with a typical bulk scalar  
such analytical exact solutions are indeed possible to obtain.

The bulk field considered in deriving these braneworld models 
is a scalar field with a non-standard coupling to gravity. In considering 
this type of scalar
field, we are surely motivated by the interesting form of the action 
for tachyon matter{\cite{sen1, sen2, sen3, tachycosmo}}, which
%The tachyon matter field  arises in the context of superstring theory
 %{\cite{sen1, sen2, sen3, tachycosmo}} and 
 finds applications in cosmology in explaining
dark matter and dark energy. 
We do not, however, restrict ourselves to the tachyon matter action as 
being exclusively string--theoretic in its origins. 
It must be noted that such an action
naturally arises in various ways -- for instance as a scalar Dirac--Born--Infeld
action or via a field theoretic generalisation of the simple 
problem of a particle falling under gravity in the presence of
a quadratic (in velocity) damping \cite{quad}.
Therefore, we would like to investigate if such matter in the bulk
could lead to interesting  braneworld models, in which the form of
the tachyon field or the tachyon potential {\em may or may not} necessarily
correspond to those obtained in specific stringy contexts. 
A second motivation behind our study is related to the fact that
one cannot fix the nature of the bulk field {\em a priori}.
There is an infinite spectrum of possibilities for constructing braneworld 
models
with different types of bulk fields and so far as they are 
physically reasonable,
well-studied in the literature and can give the right physics on the
brane, we should not hesitate in trying to figure out their consequences.
%Although physically a somewhat unusual entity,
%{\em tachyon matter}, with its non-standard coupling to gravity (compared to
%ordinary scalar fields coupled to gravity), 
Finally, the non-standard action we consider here does give rise to  
interesting physics in the context of  higher dimensional theories. 
For example, it has been shown in \cite{tachrs} that this type of field in
the bulk makes it possible to stabilise the branes (in a two--brane set--up)
which seems otherwise impossible
to achieve by the standard Goldberger-Wise mechanism \cite{goldwise}
if the backreaction of the scalar field on the metric is considered.
%String-inspired higher dimensional theories do allow  such 
%unusual scalar fields
%to reside in the bulk and they, consequently play a role, in modifying  
%physics in the real four dimensional braneworld.
With these motivating factors, 
we now move on towards constructing cosmological branes embedded
in a higher dimensional background,
arising from a non--conventional
scalar field minimally coupled to gravity 
in the bulk spacetime.
As already mentioned, the bulk geometry in this setup is warped, \textit{i.e.}, nonfactorisable.
A warped braneworld model with this type of bulk field has been proposed earlier for a flat brane 
\cite{tach}. Here we introduce a more general situation with the brane having a curvature
in the form of a cosmological (FRW) metric. 

%%%%%%%%%%%%%%%%%%%%%%%%%%%%%%%%%%%%%%%%%%%%%%%%%%%%%%%%%%%%%%%%%%%%%%%%%%%%%%%%%%%%%%%%%%

\section{The field equations with bulk matter}

The five dimensional Einstein equations with a cosmological constant and an 
arbitrary matter field in the bulk is given by
\begin{equation}
 G_{AB} = -\Lambda_5 ~g_{AB} + \kappa_5^2 ~T_{AB}^{\text{bulk}}
\label{einscalar} 
\end{equation}
The line element for the full  five dimensional warped spacetime,
with the brane representing a spatially flat, cosmological metric, is given by
the following ansatz
\begin{equation}
 ds^2 = e^{2 f(\sigma)} \left[- d t^2 + a^2(t) \left(d x^2 + d y^2 + d z^2 \right)  \right]
+ d \sigma^2 
\end{equation}
where $f(\sigma)$ is the so-called warp factor and $a(t)$ is the scale factor for the brane metric.
Here, for simplicity,  we assume  a spatially flat, cosmological brane 
though one can look into possibilities with positively/negatively curved
FRW branes.

The full five dimensional action for our braneworld model is given by 
\begin{equation}
S = S_G + S_T + S_B
\label{ac} 
\end{equation}
where the first term $S_G$ is the action for pure 5D gravity, the second term $S_T$
is that for the bulk field  and the last one  $S_B$ represents 
the contribution from the brane. Written explicitly, they read as follows \cite{rs} :
\begin{eqnarray}
S_G &=& \int d^5 x \sqrt{-g} (2 M^3 R - \Lambda_5) \label{acG} \\
S_B &=&  -\sum_{i} \lambda_i \int d^4 x \sqrt{-\gamma} \label{acB} 
\end{eqnarray} 
where $M$, $\Lambda_5$ and $\lambda_b$ are, respectively, the 5D mass scale,
the bulk cosmological constant and the brane tension
(which is a higher dimensional analogue of surface tension).
Here $g_{AB}$ is the metric for the five dimensional warped spacetime given by the coordinates
$A, B,...$ etc. and $\gamma_{\mu\nu}$ is the induced metric on the brane.

Further, the action for the bulk scalar field ($T$) is chosen to be of the form {\cite{sen1}}
\begin{equation}
S_{T} =  \int d^{5}x \sqrt{-g} V(T)\sqrt{1+g^{AB}\partial_{A} T\partial_{B} T}
\label{acT} 
\end{equation}
It should be mentioned here that
 this kind of field can simply be considered as a  scalar field non-canonically
 coupled
 to gravity, with an action of the form given in the above equation.
 Thus, in a more general sense, we are dealing with a non-standard 
 bulk field having phenomenological implications, irrespective of its origin.
Here V(T) is the potential of the field under consideration. 
Note that the bulk matter action we have written here does not have the
usual overall minus sign -- we have absorbed this sign in the potential V(T). 

The equation of motion for the field can be readily obtained  by varying the action 
w.r.t. the field T. 
This results in the following equation of motion :
\begin{equation}
\partial_{A}\left[\frac{\sqrt{-g} V(T) \partial^A T}{\sqrt{1+  (\nabla T)^2}}
 \right ] 
- \sqrt{-g} \sqrt{1+  (\nabla T)^2} \frac{\partial V(T)}{\partial T} +
\sum_i \frac{\partial \lambda_i}{\partial T} \delta (\sigma-\sigma_i) = 0
\end{equation}
Further, the bulk scalar action (\ref{acT}) when varied w.r.t. the metric gives the  
bulk energy-momentum tensor as:
\begin{equation}
T_{AB}^{\text{bulk}} =  \left[ g_{AB} V(T) \sqrt{1 +
    (\nabla T)^2} - \frac {V(T)}{\sqrt{1+ (\nabla T)^2}} 
\partial_{A} T \partial_{B} T \right] 
\end{equation}

%It should be mentioned here that
The bulk scalar field can be, in principle, a function of the extra dimension 
as well as of the spacetime (on--brane) coordinates. 
We restrict ourselves to the situation where T is of the form
$T(t,\sigma)$.
It follows from the forms of the Einstein
tensors that since there is no off--diagonal term in $G_{AB}$,  $T^{\text{bulk}}_{t\sigma}=0$. Thus,
one could have either the time-derivative or the $\sigma$-derivative of the scalar field
to vanish.  We choose to follow the first route,
\textit{i.e.}, set the time-derivative of $T$ to be zero, and obtain
the bulk Einstein equation therefrom. 
We shall show that there is, indeed, a braneworld solution for this setup where 
the brane metric has a specific (and convenient) form for the scale factor
$a(t)$ .  
One could, as well, follow the second route which may lead to qualitatively new results
arising from the time-dependence of the bulk field.

Using the ansatz for the five dimensional line element given above 
we arrive at the 
following expressions for the Einstein tensors (in the frame basis):
\begin{eqnarray}
G_{00} &=& e^{-2 f} \left( 3 \frac{\dot a^2}{a^2} \right) - 6f^{'2} - 3f^{''} 
\label{eqeintach1} \\
G_{\alpha\alpha} &=& e^{-2 f} \left( - 2\frac{\ddot a}{a} - \frac{\dot a^2}{a^2} \right) + 6f^{'2} + 3f^{''} \label{eqeintach2}\\
G_{\sigma\sigma} &=& e^{-2 f} \left(- 3\frac{\ddot a}{a} - 3 \frac{\dot a^2}{a^2} \right) + 6f^{'2} \label{eqeintach3}
\end{eqnarray}
where an overdot represents a derivative w.r.t. 
the coordinate $t$ and a prime
denotes a derivative w.r.t. the extra dimension  $\sigma$. 

The Einstein tensors given above will
have no time dependence for a de-Sitter (anti de Sitter) line element, and, 
consequently the scalar field
would be a function of  $\sigma$ alone. This choice makes the Einstein 
equations much simpler and easier to handle.
In what follows, we shall restrict ourselves to the discussion of a de-Sitter brane
for which the scale factor is given by $a(t) = e^{Ht}$. 
Consequently, the bulk Einstein equations with this type of bulk matter as the
source (and the bulk field only a function of $\sigma$) 
reduce to two coupled differential equations
\begin{eqnarray}
 3H^2  e^{-2 f}  - 6f^{'2} - 3f^{''} &=& \Lambda_5 -\kappa_5^2   ~V(T) \sqrt{1 + T^{'2}}  
 - \kappa_5^2 \sum_i ~\lambda_{i} ~\delta (\sigma -\sigma_i) \label{dstach1} \\
-6 H^2 e^{-2 f} +  6f^{'2} &=& - \Lambda_5 + \kappa_5^2  ~\frac{V(T)}{\sqrt{1 + T^{'2}}} 
\label{dstach2}
\end{eqnarray}

Given an expression for the potential $V(T)$, one can attempt to  obtain 
a set of analytical solutions for the above equations. 
Our primary aim is to solve for $f(\sigma)$ and $T(\sigma)$ resulting in 
thin or thick
brane models, for different forms of the scalar field potential.
A look at these two equations immediately suggests the following. 
We first take the $\Lambda_5$ term to the left and then 
multiply the LHS and RHS of the above equations to get the potential 
$V(T)$ by eliminating the term involving
$T'$. Similarly, we divide the LHS and RHS of the equations obtained 
after taking $\Lambda_5$ to the left and thereby obtain $T(\sigma)$.
Thus both $T(\sigma)$ and $V(T)$ are written in terms of the
warp factor and its derivatives (besides the $\Lambda_5$).
Now we choose the warp factor and hence obtain $T(\sigma)$ and
$V(T)$ explicitly.

Note, in addition, that the bulk cosmological constant $\Lambda_5$ is not an 
essential ingredient when there is matter in the bulk. 
In fact, in some cases with flat branes, the bulk scalar can play the role of 
the cosmological constant. Therefore, we shall not consider the bulk 
cosmological constant unless required.

%%%%%%%%%%%%%%%%%%%%%%%%%%%%%%%%%%%%%%%%%%%%%%%%%%%%%%%%%%%%%%%%%%%%%%%%

\section{Thin branes}

It is well-known that  one or more thin branes are realised as sharp peaks 
(derivative discontinuity) of the warp factor
at definite locations $\sigma = \sigma_i$ in the entire range of the extra 
dimension. 
To achieve this, the contribution of the brane tensions $\lambda_{i}$ (and 
the so--called four dimensional brane terms) must  be considered in the action.
The non-trivial brane action $S_B$ term, as given by Eq (\ref{acB}) 
eventually leads to the desired delta-function singularities, which 
represent the location of the branes.
Further,  we set  $\Lambda_5 = 0$, since it is not
required here to obtain the bulk geometry. 

Solving the equations (\ref{dstach1}) and (\ref{dstach2})
results in the following expressions for  the warp factor and the scalar field :
\begin{eqnarray}
f(\sigma) &=& -k |\sigma|\\
\cos {\sqrt 2} k T &=& \frac{H e^{k\vert \sigma \vert}}{k}
%T(\sigma) &=&  \frac{1}{\sqrt{2} k} \sec^{-1} \left(\frac{k}{H} e^{k |\sigma|} \right) 
\end{eqnarray}
with the potential for the scalar field to be of the form
\begin{equation}
 V(T) = \frac{3 k^2}{\kappa_5^2}  
\left[2 \sin^2(\sqrt 2 k  T ) \left( 2 - \cos^2(\sqrt 2 k T )\right) \right]^{\frac{1}{2}} 
\end{equation}
The solution for $T(\sigma)$, as given above, indicates that $\sigma$
must be bounded, i.e. $\frac{k}{H}\geq e^{k \sigma_i}$. Hence, we need
to place another brane at $\sigma=\sigma_i$ and work in the same way
as Randall--Sundrum by considering a $S^1/Z_2$ extra dimension and defining the
bulk as a slice bounded within two values of $\sigma$. The two branes
now have opposite tensions (k and -k).
In such a two-brane set-up, we must also do
a careful treatment of the bulk equation for
the tachyon field (as given above). The brane tensions need to
considered as functions of T and the discontinuity in T' (denoted as
$[T']_i$) is related
to $\frac{\partial \lambda_i}{\partial T}$ by the equation $[T']_i = -(\frac{
(1+{T'}^2)^{\frac{3}{2}}}{V(T)})_i (\frac{\partial \lambda_i}{\partial T})_i$.
We have checked that, eventually, consistency implies 
the existence of certain specific values 
of $k\sigma_0$. Choosing these values one
ensures the stability of the two-brane set-up. We do not provide the details 
of this analysis
here because we are more interested in bulk solutions in this article. 

Further, note the behaviour of the scalar field potential,
depicted in the following figure. The potential is nonsingular for any value of
$T$. In fact, it involves harmonic functions (sines and cosines). 
This oscillating behaviour is reflected in the plot as well. This property
of the potential is perhaps consistent with the nature of tachyon matter. 
However,
as already mentioned, we are not too concerned 
about the origin of the bulk scalar field,
since our main aim was to study phenomenologically inspired braneworld models.
Therefore, we do not claim that such a potential can be constructed out of
superstring theory. The point to note here is that the bulk field has some 
resemblance with tachyon matter.

\begin{figure}[htb]
\centerline{
\includegraphics[width= 7cm,height=5cm]{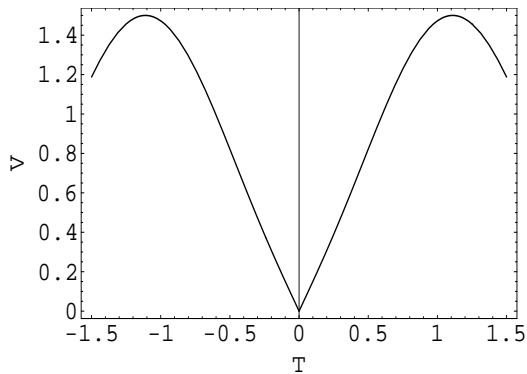}} 
\caption{The potential $V(T)$ of the bulk field $T$ resulting in a thin brane
for the values of the parameters $k = 1$ and $\kappa_{5} = 2$.}
\end{figure}

One can clearly see that here the warp factor has delta function singularities 
at the brane locations
$\sigma  = \sigma_i$. This is how we can obtain thin branes from
a bulk  field giving rise to a de-Sitter geometry on the brane.
It is noteworthy to compare the above results with the Randall-Sundrum thin brane solution \cite{rs}.
Though the above warp factor resembles RS, there are differences.
First, in the standard RS setup, the bulk is constituted of only a cosmological constant with no matter.
In our model, the bulk has a non-standard scalar field with a vanishing cosmological 
constant.
Second, and a crucial one, is that, in RS, the brane was flat. On the 
contrary, in the present model,
the induced metric on the brane represents a cosmological (de-Sitter) geometry.
However, since this is a two-brane setup, one may be curious about  the hierarchy issue,
as in the case of RS two-brane model. 
We note that since the form of the warp factor is the same as RS, 
the hierarchy problem can be solved
in this model as well. We shall, however, refrain ourselves from further details in this article. 

The variation of the  field with the extra dimension is illustrated in 
FIG. 2. 
\begin{figure}[htb]
\centerline{\includegraphics[width= 7cm,height=5cm]{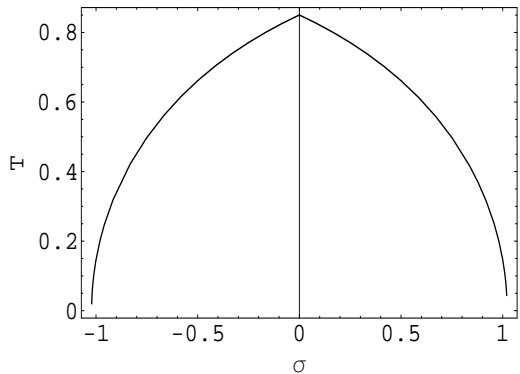}}
\caption{The  field $T(\sigma)$ as a function of the extra dimension $\sigma$
for a thin brane with decaying warp factor for $k = 1,  H = 0.36$.}
\end{figure}
The field is symmetric on the two sides of the brane (at $\sigma=0$) 
and remains nonsingular throughout the range of $\sigma$.
Further, it decays  equally from the both sides of the brane location
  as one moves away from the brane. Thus, in this setup, a decaying scalar field
gives rise to a decaying warp factor.

One might wonder whether there can be a thin brane with a growing warp factor
governed by this type of matter in the bulk. A careful look on the field equations
reveals that they have indeed a set of consistent solutions of the form
\begin{eqnarray}
f(\sigma) &=& + k |\sigma|\\
\sec {\sqrt 2} k T &=& \frac{k}{H} e^{+k\vert \sigma \vert}\\
%T(\sigma) &=&  \frac{1}{\sqrt{2} k} \sec^{-1} \left(\frac{k}{H} e^{k |\sigma|} \right) 
 V(T) &=& \frac{3 k^2}{\kappa_5^2}  
\left[2 \sin^2(\sqrt 2 k  T ) \left( 2 - \cos^2(\sqrt 2 k T )\right) \right]^{\frac{1}{2}} 
\end{eqnarray}
Here, unlike the decaying case, one can indeed construct a single
brane model with infinite extra dimension. On may also construct
models with several branes.

The above results show a change in sign in the expressions for the warp factor and the field
but the potential is unchanged.
Thus, it is indeed possible to have a growing warp factor solution
in this setup with a growing scalar field in the same potential. 

FIG. 3 depicts the nature of the field for this growing warp factor.
\begin{figure}[htb]
\centerline{\includegraphics[width= 7cm,height=5cm]{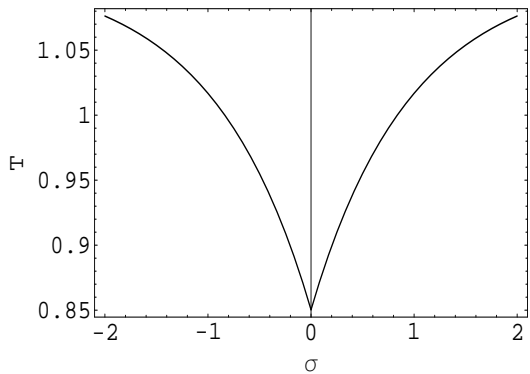}}
\caption{The  field $T(\sigma)$ as a function of the extra dimension $\sigma$
for a thin brane with growing warp factor for $k = 1,  H = 0.36$.}
\end{figure}
The field
approaches a constant value ${\sqrt 2} k T = \frac{\pi}{2}$ as one goes farther
away from the brane (on both sides), which, in turn indicates the 
asymptotically de-Sitter
(in five dimensions) nature of the bulk.
Further, here also the  field remains nonsingular throughout 
the range of $\sigma$.

Thus, we can conclude that in the thin brane scenario with a non-standard scalar field,
a decaying field under a symmetric potential gives rise to a decaying warp factor and a growing scalar field results in a  growing warp factor under the same potential.

%%%%%%%%%%%%%%%%%%%%%%%%%%%%%%%%%%%%%%%%%%%%%%%%%%%%%%%%%%%%%%%%%%%%%%%%%%%%%%

\section{Thick branes}

In order to obtain a thick brane solution with a de-Sitter geometry on the 
brane, we consider, in addition to bulk matter, a negative bulk cosmological constant.
 We shall see why we need $\Lambda_5$ here, later.
For a thick brane, we need not consider the contribution from the brane tension
because the brane is realised as a scalar field kink in the bulk.
Consequently, we shall omit the action $S_B$ for the brane part (\ref{acB}) in deriving solutions
for thick brane models.
 Hence, in this case, our intention is to solve
 the bulk field equations (\ref{dstach1}) and (\ref{dstach2}), with  $\Lambda_5 \neq 0$ but $\lambda_b =0$. 
We follow the same method as for the thin brane case, for obtaining
the solution here. 

The scalar field potential, expressed as a function of $\sigma$, is given by
\begin{eqnarray}
 V(\sigma) &=& \frac{1}{\kappa_5^2} \mbox{sech}^2(b\sigma) \sqrt{\left(\Lambda_5+ 6b^2\right)  
\sinh^2(b\sigma) + \Lambda_5 + 3 \left(b^2 -H^2\right)}  \nonumber \\
{} && \times \sqrt{\left(\Lambda_5 + 6b^2\right) \sinh^2(b\sigma) +\left(\Lambda_5 - 6H^2\right)}
\end{eqnarray}
where $b$ is some arbitrary constant. 
For this expression for the  potential,
we are able to find exact analytical expressions for $f(\sigma)$ and 
$T(\sigma)$. These are listed below.
\begin{eqnarray}
 f(\sigma) &=& \ln \cosh(b \sigma) \\
T(\sigma) &=& - \frac{i}{b} \sqrt{\frac{3(b^2 + H^2)}{\Lambda_5 - 6H^2}}
~{\text{Elliptic}} F\left[ib\sigma, \frac{\Lambda_5 + 6b^2}{\Lambda_5 -6H^2} \right] 
\end{eqnarray}

Below, we show the nature of the  field as a function of the extra 
dimension.
\begin{figure}[htb]
\centerline{\includegraphics[width= 7cm,height=5cm]{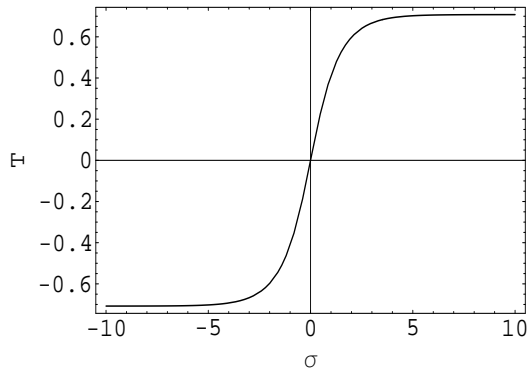}}
\caption{The bulk filed $T(\sigma)$ as a function of the extra dimension $\sigma$
for a thick brane for $\Lambda_{5} = 10$ and $b = 1 =H$.}
\end{figure}
Note that the field $T(\sigma)$
is real only if $\Lambda_5 >  6H^2$. The potential is also real for all
values of $\sigma$ is $\Lambda_5 > 6H^2$ and $b^2>H^2$. These give  constraint relation between 
the absolute value of the bulk cosmological constant, brane Hubble 
constant and the parameter $b$.
One can now easily check that for a vanishing bulk cosmological constant, the 
field $T(\sigma)$ is imaginary. This is why we have taken 
$\Lambda_5 \neq 0$.
Further, this constraint also gives rise to  real values for the 
potential for the entire range of the extra dimension.

It is worth mentioning here that the warp factor has no delta function 
singularity anywhere and the scalar field profile resembles a kink (see
figure). 
Thus, the braneworld model we obtain here represents a  thick brane.
Further, contrary to the thin brane solution, we have an increasing warp factor.
In this thick brane solution, an explicit expression for the
potential $V(T)$  as a function of the  field $T$ cannot be obtained.
However, the functional dependence of the potential on the  field 
can indeed be studied from the parametric plot of
$V(\sigma)$ versus $T(\sigma)$.
\begin{figure}[htb] 
\centerline{
\includegraphics[width= 7cm,height=5cm]{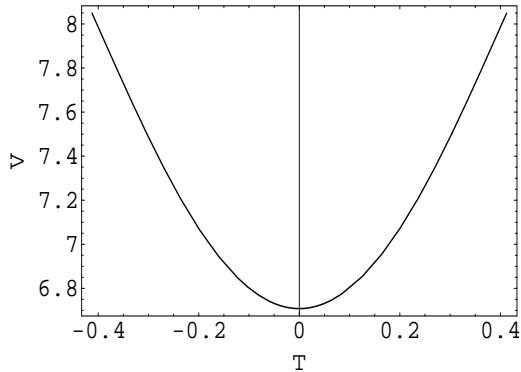}} 
\caption{The potential $V(T)$ of the bulk  field $T$ resulting in a thick brane 
with increasing warp factor with $\Lambda_{5} = 10$ and $b = 1 =H$.}
\end{figure}

From the plots, we can infer that the potential remains nonsingular for 
any real value of the  field, and, hence, throughout the domain of the extra dimension. Also, for large values of T we find that the potential
becomes asymptotically constant.

It might be asked -- what happens if $\Lambda_5 = 6H^2$ ? To get the corresponding
solution, we have to redo the calculation. It turns out, that the $T(\sigma)$
and $V(T)$ are easily obtained in this limiting case. These are:
\begin{eqnarray}
T(\sigma) &=& \frac{1}{\sqrt{2} b}\ln \vert \tanh\frac{b\sigma}{2}\vert \\
V(T) &=& -\frac{3\sqrt{2}}{\kappa_5^2} \left (b^2+H^2\right ) \frac{\cosh \sqrt{2}bT}{{\sinh^2 \sqrt{2} bT}}
\end{eqnarray}

Notice that large negative values of T correspond to $\sigma$ going to zero.
At those values the potential becomes vanishingly small. On the other hand, 
$T$ vanishes for $\sigma$ becoming infinitely large and the potential also becomes negative infinite at those points. Near $\sigma=0$, one can check that
the potential approaches $e^{-\sqrt{2} b \vert T\vert}$ which, once again, resembles the
tachyon effective potential discussed in the context of superstring theory.

Finally, we look at $\Lambda_5<6 H^2$. It is easy to see that the
potential as well as the solution become imaginary beyond a finite
$\sigma$. 

However, attempting to find a solution with a decaying warp factor
results in an imaginary value for the tachyon field at some point within
the domain of the extra dimension. So, this type of setup cannot give rise to
a thick brane with a decaying warp factor.

%%%%%%%%%%%%%%%%%%%%%%%%%%%%%%%%%%%%%%%%%%%%%%%%%%%%%%%%%%%%%%%%%%%%%%%%%%%%%%%%%

\section{Localisation of fields}

Localisation of  gravity and matter fields is an important issue in the context of
braneworld models. The basic mechanism in this endeavour is as follows:
first, we consider the concerned
field to be a function of the extra dimension as well as of the coordinates on
the brane. Then  the 4D part of the field equation is extracted out by dimensional reduction
in order to guarantee the validity of the standard field equations on the brane.
Finally, the remainder, which is a function of the extra dimension only, is subject
to the test of whether it is finite and normalisable. 
One may also check localisation of zero modes by just looking at the 5D action.
Assuming that the fields are independent of the extra dimension, one obtains
an integral over $\sigma$ whose finiteness will guarantee localisation.
%A finite value for this integral will imply
%that the field is localised on the brane.  
Thus, it is  important to address the issue of localisation of fields on the
de Sitter braneworlds arrived at in this article. 
In this section, we shall briefly sketch out this issue for different types of matter
fields as well as for gravity.
In order to do that, we  perform a conformal transformation
$d\tau = a^{-1}(t) d t$, so that the conformal time for a de Sitter brane is
given by $\tau = - H^{-1} e^{-Ht}$ and the bulk
metric can now be re-written as
\begin{equation}
d s^2 = e^{2 f(\sigma)} a^2(\tau)\left[- d \tau^2 + d x^2 + d y^2 + d z^2 \right]
+ d \sigma^2
\label{conf}  
\end{equation}
In what follows, we shall deal with this metric involving a conformally flat brane metric
for mathematical simplicity.

\subsection{Scalar field}

First, we shall address the localisation of scalar fields which are spin zero particles. 
%in the de Sitter braneworlds.
The action for a massless scalar field $\Phi$ coupled to gravity in this 5 dimensional
spacetime is given by 
\begin{equation}
S_{0} = - \frac{1}{2} \int{d^{5} x \sqrt{-g} g^{AB} \partial_{A} \Phi
  \partial_{B} \Phi} 
\end{equation}

As already mentioned, the Klein-Gordon equation in 4D curved spacetime
for this scalar field should be 
respected, so that after dimensional reduction the above action turns out to be
\begin{equation}
S_{0} = - \frac{1}{2} \int_{0}^{\infty} {e^{2f(\sigma)} d \sigma }
\int a^2(\tau) \sqrt{-\eta}{\eta^{\mu\nu} \partial_{\mu} \phi \partial_{\nu} \phi d^{4} x}
\end{equation} 
where $\phi$ is independent of $\sigma$.
The second integral readily gives the Klein-Gordon equation in  curved spacetime
on the brane. It is straightforward to check the finiteness of the integral involving the
extra dimension. 
This integral is finite for the thin brane model with decreasing warp factor
$f(\sigma) = -k |\sigma|$,
whereas it diverges away for the  increasing warp factor
$f(\sigma) = +k |\sigma|$ (thin brane) as well as for $f(\sigma) = \ln \cosh(b \sigma)$
(thick brane).
 Thus, it turns out that the zero modes of the massless scalar fields are localised on the thin brane with decreasing warp factor
whereas they are not localised on the thin or thick branes
having increasing warp factor. In particular, for the two-brane
model with a decreasing warp factor, localisation is always possible
because of the finite nature of the extra dimension.
 
\subsection{Spinor field}

The 5D action which leads to the Dirac equation in curved spacetime for
the spinor fields (spin $\frac{1}{2}$) $\Psi$ is written as
\begin{equation}
S_{\rm{Dirac}} = \int \sqrt{-g} \left(i \bar \Psi \Gamma^A D_A \Psi \right) d^5 x
\label{5Dspinor} 
\end{equation}
where $\Gamma^A$ are the 5D curved space gamma matrices which satisfy the
algebra $\Sigma^{AB} = \frac{1}{4} \left[ \Gamma^{A},\Gamma^{B}\right]$
and $D_A$ is the covariant derivative in 5D curved space. 
Considering the separation of variable to hold good, we can write
$\Psi(x^A) = \psi(x^\mu) U(\sigma)$. Substituting this back into the 5D Dirac equation
in curved space
and imposing the 4D Dirac equation in curved space on it, we obtain
the solution for $U(\sigma)$ to be $U(\sigma) = U_0 e^{-2f(\sigma)}$.
In terms of this new variable, the action (\ref{5Dspinor}) turns out to be
\begin{equation}
S_{\rm{Dirac}} = U_{0}^2 \int_{0}^{\infty} e^{-f(\sigma)} d \sigma
 \int {i a^4(\tau)\sqrt{-\eta} \bar{\psi} \gamma^{\mu} \partial_{\mu}
  \psi d^4 x} 
\end{equation}

Once again the second integral leads to the Dirac equation in curved spacetime on the brane.
So, a finite value for the first integral involving $\sigma$ will guarantee that
the zero mode of a spin $\frac{1}{2}$ field is localised on the brane. 
One can readily show that this integral diverges for the decaying warp factor
in the thin brane model as long as we have an infinite extra dimension.
However, for a finite extra dimension, the integral is finite and localisation
is possible. On the other hand, the integral 
is finite for the growing warp factors in the thin and the thick brane models
with infinite extra dimension. 
Consequently, one can conclude that the zero mode of a spinor  
is indeed localised when the de Sitter brane has an
increasing warp factor  and the extra dimension is infinite.

\subsection{Gravitational field}

Studies on localisation of gravity (spin 2) involve analysis of tensor fluctuations of
the metric under consideration. For a flat brane such as RS \cite{rs} or other flat braneworld
models, it measures the deviation from the standard inverse square law on the brane
and  accounts for its compatibility with experiments.
For a curved brane like the de Sitter brane in our model, this is generalised to the question of
whether we can get back standard cosmological results involving 4D de Sitter geometry
on the brane. 

A tensor fluctuation of the metric results in various modes for the graviton.
Our point of interest will be only on the transverse traceless modes resulting from
the metric perturbations of the 4D brane metric. 
Once again we shall use the conformal form of the metric (\ref{conf})
which will make our analysis much simpler without losing any essential information.
Thus, we have after perturbation on the brane metric :
\begin{equation}
ds^2 = e^{2f(\sigma)} a^2(\tau) \left[ \eta_{\mu\nu} + H_{\mu\nu}\right] 
 dx^{\mu}dx^{\nu} + d \sigma^2  
\end{equation} 

Using the gauge conditions $\partial_\mu H^\mu_\nu = 0$ and $H^\mu_\mu = 0$,
we assume a solution for the TT modes which satisfy the linearised wave equation in 
curved spacetime
\begin{equation}
\frac{1}{\sqrt{-g}} \partial_{M}(\sqrt{-g} g^{MN} \partial_{N} H_{\mu\nu}) = 0 
\label{waveeq} 
\end{equation}
to be of the form $H_{\mu\nu} (x^{\mu}, \sigma) = h_{\mu\nu}(x^{\mu}) 
\sum_{m} \varphi_{m}(\sigma)$.
With the use of the 4D wave equation $\Box h_{\mu\nu} = m^2 h_{\mu\nu}$ (where $\Box$
is the d'Alembertian in 4D curved spacetime),   the wave equation (\ref{waveeq}) boils down to
\begin{equation}
\varphi_m'' + 4 f'(\sigma) \varphi_m' + m^2 e^{-2f(\sigma)} \varphi_m = 0 
%\varphi_{m}'' + \left[ m^2 e^{-\frac{f(\sigma)}{2}} a^{-1}(\tau) \right] \varphi_{m} = 0 
\label{redwave} 
\end{equation} 
where $m$ is the mass of the mode under consideration and the orthonormality
condition for $\varphi_{m}$ is given by  
\begin{equation}
\int_{0}^{\infty} e^{2f(\sigma)}  \varphi_{m} \varphi_{n} d \sigma =\delta_{mn} 
\label{ortho} 
\end{equation}
For zero mode, the reduced wave equation (\ref{redwave}) does have a very simple solution
of the form $\varphi_{0} = {\rm{constant}}$, which, when substituted back into
the above orthonormality condition, will lead to the condition for localisation of
gravity on the de Sitter brane.

We can now check whether gravity is localised on the brane models proposed here.
Eq (\ref{ortho}) is satisfied by the thin brane model with decreasing warp factor
so that the zero mode gravitons are localised on this brane whereas for the 
brane models with increasing warp factor this equation is not satisfied, hence zero mode
gravitons are not localised on this brane. For a finite extra dimension
(two-brane model) localisation is always possible. 
The conclusions are similar to the case of a scalar field discussed earlier.

It is well-known that vector fields (spin 1) can not be localised in  any kind
of 5D brane models, since the integral involving the extra dimension always 
diverges for a general warp factor, irrespective of the form of the brane metric. However, the condition
is not so compelling in the case of 6D brane models. Since, in this article, we are dealing
with 5D brane models, it is obvious that the scenario will remain unaltered even
when the brane metric is cosmological, as in our case. Thus, we can say that
the vector fields are not localised on the de Sitter brane as well.

%%%%%%%%%%%%%%%%%%%%%%%%%%%%%%%%%%%%%%%%%%%%%%%%%%%%%%%%%%%%%%%%%%%%%%%%%%%%%%%%%%

\section{Summary and outlook}

In this article, we have obtained new examples of braneworlds arising 
from a non--standard scalar field in
the bulk. An important result obtained in the article is that
we can construct exact bulk models where
the induced metric on the brane represents a spatially flat
FRW geometry with a de Sitter scale factor. This, we feel, is physically
more significant than the more well--known braneworld models with 
a flat on-brane metric.
Our choice of a de-Sitter geometry on the brane is entirely 
dictated by the fact that, with it, we can arrive at exact solutions.
To this end,  we have obtained two thin brane models: one with a decreasing warp factor with two branes
and another having an increasing warp factor. We also obtain 
a thick brane model with an increasing warp factor using 
different forms for the tachyon 
fields and the potential in the bulk. The thin brane solution with a decreasing warp factor has
a warping which resembles that of RS, but now for a non-flat brane whereas we find a new
thin brane model with increasing warp factor.
The thick brane example has an 
increasing warp factor with the scalar field resembling a kink. 
The scalar field potentials are
nonsingular everywhere in the infinite domain of the extra dimension.

We have also addressed the issue of localisation of fields on the
de Sitter braneworlds arrived at in this article. 
To this end, we have analysed zero mode localisation of different spin fields
such as spin 0 (scalars), spin $\frac{1}{2}$ (spinors) and spin 2 (gravity)
by following the standard methods.
The essential conclusion from this analysis for the curved brane scenario
is that the usual results on 
field localisation (i.e. decreasing warp factors localise graviton and 
scalar fields, increasing warp factors facilitate localisation of
fermions, vector fields are not localised for either type of warping) 
go through. Thus, our model with de Sitter geometry on the brane is at par with
flat brane models, the crucial advantage being that here the 
on--brane metric is somewhat more physical since it represents 
a cosmological scenario to some extent.

In all our calculations, we have considered the scalar field to be a 
function of the extra dimension alone. Our focus was to 
demonstrate, with the help of exact solutions, that
bulk  matter can give rise to a cosmological scenario on the brane. 
These models with de Sitter geometry on the brane do have some physical 
significance.
It is well-known that a de Sitter metric represents the inflationary
scenario where there is no matter in the usual sense. Our models, therefore,
can be suitable to describe inflation from the point of view of braneworlds
where the total stress-energy tensor does not have any contribution from brane 
matter.

Of course, as in standard cosmology, matter/radiation will start forming
from reheating at the end of inflation. Hence one has to consider a non-zero 
contribution from the brane matter in this model, if one intends to describe
late time behaviour of the 4D universe. This requires the analysis
of field equations involving time-dependence of the bulk field itself.    
Thus, time-dependent solutions for the field will, surely, lead
to more realistic cosmological models. For example, restoring the 
time-dependence on the  field may result in
a spectrum of possibilities for scale factors (not necessarily
de Sitter). These may be closer to some of the
standard cosmological results, including those that represent the late-time 
(accelerating) behaviour of the universe, in a way more general than de Sitter.
In addition, on--brane line elements which are spatially non--flat (i.e
the positively and negatively curved FRW spacetimes) may also be
considered, for which, one can search for consistent solutions with 
bulk  matter. In either of the above--mentioned possibilities
one has to deal with much more complicated equations and we guess
it to be a highly nontrivial exercise possibly requiring extensive
numerical computation. We hope to address these issues in the near
future.  

%%%%%%%%%%%%%%%%%%%%%%%%%%%%%%%%%%%%%%%%%%%%%%%%%%%%%%%%%%%%%%%%%%%%%%%%%%%%%%%%%%%

\end{document}